\documentstyle[12pt]{article}

\thicklines                         
\def\co{{\cal O}}

\def\a{\alpha}
\def\b{\beta}

\def\d{\delta}
\def\e{\epsilon}                
\def\f{\phi}                    

\def\h{\eta}

\def\j{\psi}

\def\l{\lambda}
\def\m{\mu}
\def\n{\nu}

\def\p{\pi}                     
\def\r{\rho}                    
\def\s{\sigma}                  

\def\x{\xi}

\def\F{\Phi}

\def\J{\Psi}
\def\L{\Lambda}
\def\O{\Omega}

\def\svev#1{\left\langle #1\right\rangle}             
\def\boxx{\raisebox{-.4ex}{\large$\Box$}}             
\def\sbra#1{\left\langle #1\right|}             
\def\sket#1{\left| #1\right\rangle}             
\def\Tilde#1{\widetilde{#1}}                    

\def\NON{\nonumber\\}

\def\rhs{\mbox{r.h.s.} }
\def\lhs{\mbox{l.h.s.} }
\def\ie{\mbox{i.e.} }
\def\eg{\mbox{e.g.} }

\def\beq{\begin{equation}}
\def\eeq{\end{equation}}
\def\bqry{\begin{eqnarray}}
\def\eqry{\end{eqnarray}}
\def\beqrabc#1{ \setcounter{equation}{0}
                \renewcommand{\theequation}{#1\alph{equation}}
                \begin{eqnarray} }
\def\secteq#1{ \setcounter{equation}{0}
               \renewcommand{\theequation}{#1.\arabic{equation}} }

\def\seeq#1{eq.~(\ref{#1})}
\def\seEq#1{Eq.~(\ref{#1})}
\def\seeqs#1{eqs.~(\ref{#1})}

\def\seneq#1{~(\ref{#1})}

\def\JMP#1{Jour. Math. Phys. {\bf #1}}
\def\NPB#1{Nucl. Phys. {\bf B#1}}

\def\PLB#1{Phys. Lett. {\bf B#1}}
\def\PRD#1{Phys. Rev. {\bf D#1}}

\def\PRP#1{Phys. Rep. {\bf #1}}

\renewcommand{\thefootnote}{\fnsymbol{footnote}}

\begin{document}
\hyphenation{fer-mio-nic per-tur-ba-tive}

\noindent June 1995 \hfill TAUP--2253--95

\hfill hep-th/9506051
\par
\begin{center}
\vspace{15mm}
{\large\bf Anomalous Supersymmetry Breaking by Instantons}\\[10mm]
A.\ Casher\footnote{Email: ronyc@ccsg.tau.ac.il},
V.\ Elkonin\footnote{Email: elkonin@ccsg.tau.ac.il} and
Y.\ Shamir\footnote{Email: ftshamir@wicc.weizmann.ac.il}\footnote{
Work supported in part by the US-Israel Binational Science
Foundation, and the Israel Academy of Science.}\\[5mm]
{\it School of Physics and Astronomy\\
Beverly and Raymond Sackler Faculty of Exact Sciences\\
Tel-Aviv University, Ramat Aviv 69978, ISRAEL}\\[15mm]
{ABSTRACT}\\[2mm]
\end{center}

\begin{quotation}

  We show that instantons violate a supersymmetric identity
in a classically supersymmetric Higgs model with no massless fermions.
This anomalous breaking arises because the correct perturbative expansion
in the instanton sector is not supersymmetric. The attempt to construct a
manifestly supersymmetric expansion generates infra-red divergences.

\end{quotation}

\setcounter{footnote}{0}
\renewcommand{\thefootnote}{\arabic{footnote}}


\newpage
\noindent {\large\bf 1.~~Introduction}
\vspace{3ex}

  Asymptotically free supersymmetric gauge theories may play an important
role at very short distance scales.
Present day understanding of the physical properties of these
theories~[1-6] relies in a crucial way on the {\it conjecture} that
non-perturbative effects do not break supersymmetry (SUSY) explicitly.
The plausibility of this conjecture has been discussed in the literature.
On the other hand, there are also arguments against it~\cite{cs1,cs2}.

  In this paper we show that the one instanton contribution to a certain
condensate violates the relevant supersymmetric identity. We consider a
SUSY-Higgs model, which allows for a fully controlled
calculation of non-perturbative effects. The model contains no massless
fermions. Hence, the option of spontaneous ``dynamical'' SUSY breaking
is ruled out.

  Anomalous SUSY breaking is closely related to the Higgs mechanism,
which provides a physical infra-red cutoff through the generation of masses.
Physical observables receive non-supersymmetric contributions
which arise from the non-constancy of the Higgs field in the instanton
sector.

  A {\it formally} supersymmetric expansion can be constructed in the
instanton sector if the effects of the classical Higgs field, in particular
the generation of masses, are incorporated {\it perturbatively}.
Only the background gauge field is taken into account at tree level.
In this formal expansion one should find supersymmetric
results {\it provided the results are finite}.

  The price paid for the perturbative treatment of mass terms is
the appearance of  infra-red (IR) divergences.
In the literature on supersymmetric QCD there are several leading
order instanton calculations with a non-vanishing scalar VEV~\cite{ads,itep}.
In all of them, IR divergences will arise at the next-to-leading order.
In the case of the condensate calculated below, there are IR divergences
already at the leading order. Moreover, the IR divergences become more
and more severe as the order increases.

  The anomalous breaking of SUSY is triggered by these IR divergences.
While the supersymmetric calculus is valid for some (but not all) leading
order calculations, the supersymmetric Feynman rules are in general
ill-defined. As a result, an IR cutoff must be introduced explicitly.
But the IR cutoff necessarily breaks SUSY if we are interested in the one
instanton sector\footnote{
  We comment that periodic boundary conditions or the compactification of
$R^4$ to $S^4$ are both incompatible with the asymptotic behaviour of the
Higgs field, which reflects the topology of the instanton sector.
}.
This sets the stage for an anomaly of a new kind.
As we discuss in detail below, the IR divergences are eliminated by a
resummation procedure that reconstructs massive propagators from
massless ones. This resummation gives rise in general
to non-supersymmetric results, because the spectrum of massive modes in the
instanton sector is not supersymmetric.

  Let us describe the massless supersymmetric expansion in more
detail. As the background gauge field one takes the pure
instanton expression. (The source current in the gauge
field's equation of motion is neglected to leading order).
For simplicity we will henceforth assume that all masses arise through
the Higgs mechanism. Mass insertions are treated perturbatively.
Thus, zero modes and propagators are obtained by solving
the massless field equations in the pure instanton background.
The classical Higgs field itself
is a solution of the equation $D^2 \f =0$, which satisfies the boundary
condition $\f(x)\to v$ at infinity. In this framework, supersymmetric
relations exist at two levels. As shown in ref.~\cite{itep} the classical
fields and the fermionic zero modes are related via SUSY transformations.
Also, supersymmetric relations exist between the propagators, because the
continuous spectrum in the pure instanton background is universal~\cite{t}.

  The supersymmetric expansion reflects an approximation which is
valid in the instanton's core. In a Higgs model, the leading order
effective action in the instanton sector is~\cite{t}
\beq
  S_E(\r) = 8\p^2/g^2(\r) + 4\p^2 v^2 \r^2 \,.
\label{seff}
\eeq
Here $\r$ is the instanton's size. The numerical coefficient of the second
term, which represents the contribution of the Higgs field(s),
is valid for an SU(2) gauge group~\cite{ads}.

  \seEq{seff} implies that the saddle point of the
$\r$-integration occurs at $\r\sim v^{-1}$. In units of the Higgs VEV,
the gauge field's strength in the core is $O(1/g)$.
In contrast, the magnitude of the Higgs field is O(1), namely it is
comparable to a typical quantum fluctuation on the same scale.
This provides a formal justification for keeping only the gauge field at
tree level. Propagators, insertions of the classical
Higgs field and the full set of zero modes of the pure instanton background
are all treated as elements for the construction of Feynman graphs.

  The IR divergences arise  from multiple mass insertions on a given line in a
Feynman graph. Their physical origin is easily understood.
Consider the linearized field equations taking into account {\it all}
classical fields in the usual way. These differential equations determine
the exact form of propagators and zero modes.
In the core, the equations are dominated by the covariant
derivative terms. Consequently, the supersymmetric expressions provide
good first approximations for the exact ones.

  In the transition region $v^{-1} \ll |x| \ll m^{-1}$ the equations are
dominated by the free kinetic term. (Here $m$ stands for a generic
Higgs-induced mass). However, for $|x| \sim m^{-1}$ and at larger
distances, the mass terms are comparable to the kinetic term.
At any distance scale which is large compared
to the instanton's size, the massless field equations
give rise to a power law behaviour.  But since the relevant fields are
{\it massive} the correct asymptotic behaviour is a falling exponential.
Within the supersymmetric Feynman rules, one must sum over an arbitrary
number of mass insertions on every line in order to reproduce
the falling exponential. The sum of a finite number of mass insertions
will in general be IR divergent.

  The IR divergences of the massless supersymmetric expansion are
therefore spurious, in the sense that they arise from using expressions that do
not have the correct asymptotic behaviour. This statement applies to all
elements of the perturbative expansion, namely, to propagators, to zero modes
and to the deviations of the classical fields themselves from their limiting
values.

  The massive Feynman rules in the instanton sector are obtained by applying
the standard variational principle.
To make contact with the massless supersymmetric expansion,
it is convenient to separate out the integration over the instanton's size.
This is done by first fixing the
size by introducing a constraint~\cite{constr}. For any value of the
constraint parameter one sets up a perturbative expansion in a
standard manner. In the end, one integrates over the instanton's size which is
now represented by the constraint parameter $\r$.
A judicious choice of the constraint will modify mainly the classical gauge
field's equation. Physically, the source current arising from the matter part
of the lagrangian tends to shrink the instanton's size, and this effect
is balanced by the contribution of the constraint.

  The classical fields obtained via this procedure tend exponentially
to their vacuum values at large distances. The exponential asymptotic
behaviour also characterizes the massive propagators and the surviving
exact zero modes. As in scattering theory,
this allows the removal of the finite volume cutoff, and one can work
directly in the infinite volume limit.

  The massive Feynman rules feature a supersymmetric set of vertices, because
the classical lagrangian is supersymmetric. But in general there are no
supersymmetric relations between the classical fields and the zero modes' wave
functions\footnote{
  Models with $N_f=1$ are a special case. An example is the simpler Higgs
model obtained by dropping the ``lepton'' families from the model of Sect.~2.
In this case, supersymmetric relations between the classical
fields and the zero modes survive to some
extent also within the massive Feynman
rules. This is a reminiscent of the situation in less than four dimensions,
where the topologically non-trivial objects are exact solutions of
the classical field equations. In both cases the SUSY violation arises from
the absence of local supersymmetric relations betweens the spectra of
bosonic and fermionic fluctuations. As a result, however, there are no
SUSY violations at the level of tree diagrams.
See ref.~\cite{cs1} for more details.
}.
Moreover, the massive fluctuations spectrum is no longer supersymmetric
(except approximately at very short distances).
The linearized bosonic field equation and the square of the Dirac
operator define two Schr\"odinger-like operators.
Due to the non-constancy of the Higgs field in the instanton's core,
these Schr\"odinger operators involve {\it different}
potentials. Consequently, massive bosonic and
fermionic eigenstates are not related by the action
of a local differential operator~\cite{cs1}.
Therefore, the breaking of SUSY is ultimately attributed to the
non-invariance of the path integral measure.

  We summarize the relation between the massive Feynman rules and the massless
ones by considering the example of an exact zero mode, \ie a zero
mode that survives the introduction of the Higgs field. Other elements
of the Feynman rules (propagators etc.) exhibit analogous behaviour.
The \lhs of any field equation can be written as $H_0 + V$, where $V$
contains the dependence on the Higgs field and $H_0$ contains all the rest.
Let $\J_0$ be a zero mode of $H_0$ and $\J$ the corresponding zero mode
of $H_0 + V$. Formally, $V$ is small compared to $H_0$, and so we
may think of reexpanding $\J$ using the Born series
$\J = \J_0 - G_0 V \J_0 + \cdots$. But in general this is not possible
because $V$ contains a mass term that changes the asymptotic behaviour.
Individual terms in the Born series have divergent norms, and
their insertion in a Feynman graph gives rise to IR divergences.

  In practice, one has to distinguish between two cases.
Suppose that the massless Feynman rules give
rise to a finite answer for some observable (or, more generally, for some
sub-diagram) at the leading order.
As far as a given zero mode is  concerned, this will typically
happen when the leading order result involves only
the ``zeroth approximation'' $\J_0$ which is obviously normalizable.
Notice that the difference $\J-\J_0$ is normalizable too.
Replacing $\J_0$ by $\J-\J_0$ will measure the difference between the
correct result and the (still finite) prediction of the massless expansion.
One can show that, up to logarithmic corrections,
the contribution of the difference $\J-\J_0$ will be
damped by two powers of the coupling constant(s).
A similar pattern is found for other elements
of the Feynman graphs. In conclusion, if the massless supersymmetric
rules give rise to a finite answer for a (sub-)graph at the leading order,
then the answer is valid at that order, and it will be reproduced by the
massive Feynman rules.

  Whenever the massless supersymmetric rules give rise to IR-divergent
expressions, the massive Feynman rules must be invoked. The mass terms then
provide a {\it physical} cutoff for the formally IR-divergent integrals.
Below we encounter a sub-graph which is equal, within the massless
supersymmetric rules, to the norm squared of the first Born correction
$\J_1 = - G_0 V \J_0$ to a certain zero mode.
However, the norm of $\J_1$ is IR-divergent.

  Physically, $\J_1$ represents a new field component
that the original zero mode develops due to the mass terms.
Using the massive Feynman rules, what has to be
calculated is the norm of the new component. (The overall
normalization of the zero mode is still determined by the original component).
The new component's norm has a leading logarithmic piece which is
easily computed, and which contributes to the final result \seeq{f0}.

  The IR divergences signal the onset of a {\it non-analytic} dependence on
Higgs masses, and, hence, on coupling constants.
The condensate calculated below features a
non-analytic dependence on two Yukawa couplings. This result is of course
beyond the scope of the massless supersymmetric expansion, which contains only
positive powers of the Yukawa couplings. More generally, we expect the
appearance of analogous non-analytic dependence on the {\it gauge coupling}
too. Technically, disentangling non-analytic short distance effects (the
RG flow) from long  distances ones is more complicated in this case.

  This paper is organized as follows. In Sect.~2 we define the SUSY-Higgs
model. In Sect.~3 we present our results. Sect.~4 contains our conclusions.
Some technical detail are relegated to an Appendix.

\vspace{5ex}
\noindent {\large\bf 2.~~The model}
\vspace{3ex}

 The charged fields of our
SU(2)-Higgs model are the same as in supersymmetric QCD with $N_c=N_f=2$.
Two charged doublets play the role of Higgs superfields. The other pair of
charged superfields together with a pair of neutral ones make
two supersymmetric ``lepton'' families.
  The Higgs superpotential is
\beq
  W_1 = h \F_0 \left(
       {1\over 2} \e_{ij} \e_{AB}\, \F_{iA} \F_{jB} - v^{2}
       \right)\,.
\label{sp}
\eeq
Here $\F_0=(\f_0,\j_0)$ is a neutral chiral superfield.
$\F_{iA}=(\f_{iA},\j_{iA})$ contain the two Higgs
doublets and their fermionic partners. The indices $A,i=1,2$ correspond to
SU(2) colour and flavour groups respectively. (The flavour SU(2)
plays the role of SU(2)$_R$ in the analogy to the
supersymmetric Standard Model). We use the following representations
$T^a_{AB} = -{1\over 2} \s^a_{BA}$ and $F^a_{ij} = {1\over 2} \s^a_{ij}$
for the colour and flavour generators respectively.

  The two extra charged superfield are denoted $\h_{A\pm}$, and the two
extra neutral ones are $\x_{i\pm}$.  These letters will also be used to denote
the fermionic components. The scalar components are denoted
$\tilde\h_{A\pm}$ and $\tilde\x_{i\pm}$.
The $\pm$ index labels the two ``lepton'' families.
Notice that these families form a doublet under a horizontal SU(2).
The full superpotential is
\beq
  W=W_1+W_2 \,,
\eeq
\newpage
\noindent where
\bqry
  W_2 & = & y\, \e_{ij} \e_{AB}\, \F_{jB}
        \left( \x_{i+}\, \h_{A-} - \x_{i-}\, \h_{A+}  \right) \NON
      & & +\,  m_0\, \e_{ij}\, \x_{i+} \x_{j-} \,.
\label{sp2}
\eqry
The mass parameter $m_0$ will be treated as a small
perturbation\footnote{
The perturbative treatment of $m_0$ does not generate
any IR divergences because all the fields are already massive for $m_0=0$.
Our results can also be regarded as a calculation of the
integrated form of the \rhs of \seeq{m0} in the model with $m_0=0$.
}
and we will work to first order in $m_0$.

  The classical potential has a unique supersymmetric minimum
(up to colour and flavour transformations).
The only non-vanishing VEV is $\svev{\f_{iA}} = v \d_{iA}$.
This minimum breaks the gauge symmetry completely, but it leaves unbroken
the diagonal SU(2) generated by $T^a + F^a$. Under this vector SU(2), the
two Higgs superfields decompose into a singlet
$\F' = \d_{iA} \F_{iA}/\sqrt{2}$
and a triplet $\F^a = \s^a_{iA} \F_{iA}/\sqrt{2}$.

  All fields acquire masses through the Higgs mechanism. In the triplet sector
(which includes the gauge and the $\F^a$ supermultiplets) the mass is $\m=gv$.
The mass of the singlet fields
$\F'$ and $\F_0$ is $m=\sqrt{2}hv$. For $m_0=0$, the mass of the lepton
families is $m_1 = yv$.

  The model has two approximate $R$-symmetries. U(1)$_R$ is a non-anomalous
symmetry, which becomes exact in the limit $m_0=0$. The other one, denoted
U(1)$_X$, is a classical symmetry of the full lagrangian, but it is
broken explicitly by instantons.  The fermion charges under
the $R$-symmetries are given in Table~1 in units of the gaugino's
charge. As usual, the charges of the corresponding
scalars are related by $Q_R(scalar) = Q_R(fermion)+1$.


\vspace{8ex}

%
%
\begin{table}[hbt]
\begin{center}
\begin{tabular}{|c|c c c c|}   \hline
 & $\j_{iA}$ & $\j_0$ & $\h_{A\pm}$ & $\x_{i\pm}$ \\ \hline \hline
U(1)$_R$  & -1 & 1 & -1 & 1 \\ \hline
U(1)$_X$  & -1 & 1 & 0 & 0 \\ \hline
\end{tabular}
\vspace{1ex}
\caption{Fermion charges under the $R$-symmetries}
\vspace{1.0cm}
\end{center}
\end{table}


\newpage
\noindent {\large\bf 3.~~Results}

\vspace{2ex}
\noindent {\bf 3.1~~The SUSY identity}
\vspace{1ex}

  Let $\co(x)$ denote the lowest component of a gauge invariant chiral
superfield and let $\bar{Q}$ be a SUSY generator. Using $[\bar{Q},\co(x)]=0$
one arrives at the following on-shell SUSY identity
\bqry
  0 & = & \svev{ \co(z)\, \{ \bar{Q},\,
        \tilde\x^*_{i+} \bar\x_{j-}(x) \} } \NON
    & = & \svev{ \co(z) \left( (1/2)\, \bar\x_{i+} \bar\x_{j-}(x) +
        y\e_{jk}\e_{AB}\,
        \tilde\x^*_{i+} \f_{kA} \tilde\h_{B+}(x) \right) } \,.
\label{m0}
\eqry
This identity is related to the analytic properties of the condensate
$\svev{\co}=\svev{\co(z)}$ as follows.
Let us assume that $m_0$ in \seeq{sp2} is a complex parameter.
On the second row of \seeq{m0} we contract with $\e_{ij}$ and
integrate over $x$. A similar identity is obtained by the interchange of
$\x_+ \leftrightarrow \x_-$ and $\h_+ \leftrightarrow -\h_-$.
Taking into account the two identities we arrive at the variation of
$\svev{\co}$ with respect to $m_0^*$. Therefore, \seeq{m0}
requires that $\svev{\co}$ be a function of $m_0$ only,
but not of $m_0^*$~\cite{it}.

\vspace{2ex}
\noindent {\bf 3.2~~Calculation of $\svev{\f_0}$}
\vspace{1ex}

  Our main result is that the SUSY identity \seeq{m0} is violated for
$\co=\f_0$. An explicit calculation gives rise to
\beq
  \svev{\f_0} = m_0^* \, {\L^4\over v^4} {y^2\over 4\p^2 g^4 h} \log y
                + \cdots \,.
\label{f0}
\eeq
The dots stand for subleading terms\footnote{
The anti-instanton sector makes a contribution to
$\svev{\f^*_0}$ that satisfies the on-shell relation
$\svev{\f^*_0}=\svev{\f_0}^*$. We comment that $\svev{\f_0}$ vanishes
to all orders in perturbation theory because $\f_0$ carries a non-zero
U(1)$_X$ charge.
}.
$\L$ is the one-loop RG invariant
scale of the theory. We expect similar violations for
other condensates, such as for example $\svev{\l\l}$. However, for reasons
that we explain below the calculation of the $m_0^*$-dependence of
$\svev{\l\l}$ is more complicated.

  In the calculation of $\svev{\f_0}$ we
adopt the following strategy. Diagrams will first be drawn using the massless
supersymmetric rules, and we will try to apply these rules to evaluate every
sub-graph. In view of the relation between
the massive Feynman rules and the massless ones,
as discussed in the introduction,
the result for a given sub-graph is valid {\it as long as it is finite}.
When the massless rules produce IR divergent expressions,
the massive Feynman rules will be invoked to calculate the relevant integrals.

  The diagrams  that contribute to the leading order result \seeq{f0} are
depicted in Figs.~1 and~2. The two diagrams correspond to the two terms on the
last row of \seeq{m0}.   The vertex marked with a thick cross
is linear in $m_0^*$. The dotted lines that end with a cross represent
insertions of the classical Higgs field. They differ from ordinary mass
insertions by the presence of short range potential terms that arise from
the non-constancy of the Higgs field in the instanton's core.

  Each thick line that emanates from the instanton (the shaded circle)
represents one of the eight zero modes that the model had had in the
absence of the Higgs field. These include the four gaugino zero modes and one
zero mode for each charged doublet. With the classical Higgs field turned on,
only the $\l^{SS}$ pair and the $\h_{\pm}$ pair remain  exact zero modes
(with modified wave functions that contain new field components).
The Higgsino and $\l^{SC}$ zero modes mix through the Higgs field.
However, they are still {\it approximate} zero modes because the Higgs field
is a small perturbation over their support. More precisely,
the $\l^{SC}$ and Higgsino zero modes become either resonances or
bound states with non-zero eigenvalues.
(The former possibility is more likely). This is the physical
justification for treating their mixing using the massless Feynman rules,
with the usual reservation that this does not give rise to IR divergences.
Further details on both the exact and the approximate zero modes are given
in the Appendix.

  Before we continue with the calculation of Figs.~1 and~2 let us discuss
the role of other diagrams. If we only count powers of coupling constants that
come from the vertices and ignore the IR divergences momentarily,
we should consider diagrams which are $O(g^4 y^2 h)$. Now, within
the massless Feynman rules, all these diagrams have in common a quadratic IR
divergence due to the external $\f_0$ leg. If we try to use
a (free) massless propagator instead of the correct $\f_0$ propagator,
we encounter at some point an integral of the form
$\int d^4 x / x^2$. As we discuss below, within the
massive Feynman rules the quadratic divergence boils down to a factor
of $1/m^2$. We then use $m^2=2 h^2 v^2$ to separate out a dimensionful
factor of $v^2$. The remaining $1/h^2$, together with the explicit
factor of $h$ coming from
the vertex, gives rise to the $1/h$ factor in \seeq{f0}.

  Apart from the common quadratic IR divergence, there are several diagrams
that contain an extra logarithmic IR divergence of the form
$\int d^4x/(x^2+\r^2)^2$.  The physical IR cutoff of such an integral is
an inverse mass scale. Thus, the integral behaves like the
logarithm of the relevant Higgs induced mass times $\r$. In view of
the relation $\r\sim v^{-1}$, this is equivalent to the logarithm
of a coupling constant.

  An enhancement factor of $\log y$ arises if and only if the logarithmic
divergence comes from the ``lepton'' sector, the relevant diagrams being
Figs.~1 and~2. As our explicit calculation shows, the leading logs do not
cancel between the two diagrams, which sum up to the final result
\seeq{f0}.

  Notice that we have indicated in the figures which insertion of the Higgs
field corresponds to $\f_{iA}$ and which to its complex conjugate. Other
diagrams that are formally of the same order can be obtained by replacing one
insertion of the Higgs field and one insertion of its complex conjugate
by a $\svev{\f\,\f^*}$ propagator. However, the resulting diagrams no longer
have the extra logarithmic divergence, because the propagators provide
extra powers of $1/x^2$ (some examples are discussed in part III of the
Appendix). The leading logs arise only from diagrams with a maximal number
of insertions of the Higgs field.

  Within the massless Feynman rules, there are also individual diagrams
containing logarithmic IR divergences  related to the $\l^{SC}$ and
Higgsino zero modes. But now the offensive $1/(x^2+\r^2)^2$
behaviour cancels out between different diagrams {\it before}
the $\r$ integration. This implies that the corresponding diagrams of the
{\it massive} Feynman rules contain no logarithmic factors\footnote{
  Had the logarithmic divergences coming from the gaugino-Higgsino
sector {\it not} cancelled out, they would have represented an independent
contribution to $\svev{\f_0}$, because they could not give rise to $\log y$,
but only to a linear combination of $\log g$ and $\log h$.
}.
The physical reason for this behaviour is the following. A slow $1/x^2$
decrease of a zero mode's wave function at intermediate distances can arise
only from new field components not present in the original zero mode.
The new subleading field components are well defined in the case of an
{\it exact} zero mode. But subleading field components of {\it approximate}
zero modes cannot be disentangled from the continuous spectrum. For further
details see part III of the Appendix.

  We now return to the derivation of \seeq{f0}.
The diagrams are evaluated with the measure (see \eg ref.~\cite{it})
\beq
  2^9 \p^6 \L^4 g^{-8} \int d\r^2 \r^2 e^{-4\p^2 v^2\r^2} \ldots
\label{msr}
\eeq
For Fig.~1 the integrand of the $\r$-integral is
\beq
  \left( {g^2 v^2 \over 2} \right)
  \left( {g^2 \r^2 \over 8h} \right)
  \left( m_0^* (yv\r)^2 \log y  \right) \,.
\label{intg}
\eeq
Let us explain this expression. Consider first the mixing of
the $\l^{SC}$ and Higgsino zero modes~\cite{ads} through the
Yukawa-gauge coupling
$ig\sqrt{2}\, \l^a \f^*_A T^a_{AB} \j_B$ . To leading order there are
no IR divergences, and one can calculate the mixing using formal first
order perturbation theory. Substituting the pure instanton expressions for
the zero modes \seeqs{lsc} and\seneq{higgsino}, and for the Higgs
field \seeqs{hf} and\seneq{fx}, we arrive at
the first factor in \seeq{intg}.

  The second factor corresponds to the $\f_0$ tadpole,
namely to the $\f_0$ line  that emanates from the pair of $\l^{SS}$ zero
modes (see Fig.~3). It is convenient to trade the integration
over the instanton's collective coordinates with an integration over the
external point $z$, keeping the instanton at the origin. The tadpole is
\beq
  {h \over 2} \e_{kl}\e_{ij}\e_{AB}\, \int d^4x\, d^4z\,
  \Tilde{G}(z,x)\, \bar\j^k_{iA}(x) \bar\j^l_{jB}(x) \,.
\label{bottom}
\eeq
Here $\Tilde{G}(z,x)$ is the $\f_0$ propagator and $\bar\j^k_{iA}(x)$ is a
new field component of the $k$-th $\l^{SS}$ zero mode ($k=1,2$).
We immediately see that the attempt to
treat $\f_0$ as a massless field gives rise to a quadratic
IR divergence. Within the massive Feynman rules,
the $\f_0$ propagator is defined by
\beq
  (-\boxx_z + m^2 + U) \Tilde{G}(z,x) = \d^4(z-x)\,.
\eeq
The short range potential $U$ is a function of the radial coordinate $r=|z|$
only. Explicitly $U(r)=m^2(\varphi^2(r)-1)$. (See \seeq{hf} for the
definition of $\varphi(r)$). The short range potential {\it can} be neglected
to leading order. We therefore substitute a free massive propagator for
$\Tilde{G}(z,x)$ and find
\beq
  \int d^4z\, \Tilde{G}(z,x) = {1\over m^2} \,.
\label{intprop}
\eeq
Notice that the result is independent of $x$.

  The rest of \seeq{bottom} does not contain any divergent factors.
Instead of using the exact wave function of the $\bar\j(x)$
component of the $\l^{SS}$ zero mode, we can apply
first order massless perturbation theory. The general formula for the
first Born correction is $\J_1=-G_0 V \J_0$. In the
present case $G_0$ stands for the $\svev{\j\,\bar\j}_0$ propagator
in the pure instanton background, $V$ stands for the Yukawa-gauge coupling
as a function of the Higgs field,
$\J_0$ stands for the original $\l^{SS}$
zero mode \seeq{lss} and $\J_1$ stands for the new $\bar\j(x)$ field component.
In practice, instead of finding $\bar\j(x)$ by integration,
it is easier to solve the corresponding differential equation
$H_0\J_1= - V \J_0$, where $H_0$ is the massless covariant Dirac
operator. The result is
\beq
  \bar\j^k_{iA\a}(x)=
  {igv\over 2\p} {\e_{i\a}\d_{Ak}\,\r^2\over (x^2+\r^2)^{3\over 2} }\,.
\label{newj}
\eeq
Here $\a$ is the spinor index. Carrying out the $x$-integration
and using $m^2=2h^2v^2$ we arrive at the second factor
in \seeq{intg}.

  The last factor in \seeq{intg} corresponds to the upper part of Fig.~1.
It takes the form
\beq
  m_0^*\,\e_{ij} \int d^4x\, \bar\x_{i+}(x) \bar\x_{j-}(x) \,,
\label{intx}
\eeq
where $\bar\x_\pm(x)$ is a new field component of the $\h_\mp$ zero mode.
As before, we try to use the first Born correction. The result is
\beq
  \bar\x_{i\a\pm}(x) = \mp {yv\over 2\p} {\e_{i\a}\,\r\over x^2+\r^2}\,.
\label{approx}
\eeq
In this case the spacetime integration is logarithmically IR divergent.
The true form of $\bar\x_\pm(x)$ is determined
by eq.~(A.12). It agrees with
\seeq{approx} for $|x|\ll m_1^{-1}$, whereas
for $|x|\gg m_1^{-1}$, $\bar\x_\pm(x)$ becomes a falling exponential.
(That the transition occurs at the distance scale $m_1^{-1}$
is a general feature of massive linear equations. See part II of the
Appendix for more details).

  Let us write $d^4x= r^3\,dr\,d\O$, and
split the radial integration in \seeq{intx} into three regions
\beq
  \int_0^\infty = \int_0^\r + \int_\r^{m_1^{-1}}
		  +\int_{m_1^{-1}}^\infty \,.
\label{split}
\eeq
The integral is dominated by the intermediate region, where one has
$\bar\x_{+} \bar\x_{-}(x) \sim 1/x^4$, leading to
\beq
  \int_\r^{m_1^{-1}} {d^4x\over x^4} = -2\p^2 \log(m_1\r) \,.
\eeq
Notice that $\log(m_1\r)=\log y + \log(v\r)$, and that $\log(v\r)$ is $O(1)$.
Neglecting $O(1)$ contributions, we complete the calculation by adding
the appropriate prefactors required by \seeq{approx}. The result is the
last factor in \seeq{intg}.

  The calculation of Fig.~2 is similar. Using massless perturbation
theory we find the ``induced scalar field''
\beq
  \tilde\h^k_{A\pm}(x) = - {ig\over 4\sqrt{2}\, \p^2}
               {\e_{kl} x^\m \bar\s^\m_{lA} \over (x^2+\r^2)^{3\over 2} }\,.
\label{indh}
\eeq
This induced field is obtained by solving the massless field equation with
a source given by the
product of the $k$-th $\l^{SC}$ zero mode and an $\h$ zero mode.
The second induced field, whose source is the product of the $n$-th
Higgsino zero mode ($n=1,2$) and the other $\h$ zero mode, is
\beq
  \tilde\x^{n*}_{i\pm}(x) = \pm {y\over 4 \p^2} {\e_{in}\over x^2+\r^2} \,.
\label{indx}
\eeq
Now we have a logarithmic IR divergence from the integral
\beq
  m_0^*\, y\, \e_{AB} \int d^4x\, \tilde\x^{n*}_{i\pm}(x)
  \f_{iA}(x) \tilde\h^k_{B\pm}(x) \,.
\label{bosons}
\eeq
As before, the leading log is obtained by substituting \seeqs{indh}
and\seneq{indx} in \seeq{bosons} and imposing a physical cutoff at
$r_0 \sim m_1^{-1}$.  Summing the leading logs from the two diagrams we find
\beq
  \svev{\f_0} = m_0^* \L^4 v^2 {16\p^4 y^2\over g^4 h} \log y
  \int d\r^2\, \r^4 e^{-4 \p^2 v^2\r^2} (2\p^2 v^2\r^2 - 1) \,.
\label{sum}
\eeq
In this equation, the first and second terms in parenthesis correspond
to Figs.~1 and~2 respectively. Carrying out the $\r$-integration
we finally arrive at \seeq{f0}. Further discussion of the physical mechanism
underlying this result is given in Sect.~4.

\vspace{2ex}
\noindent {\bf 3.3~Comparison of $\svev{\f_0}$ and $\svev{\l\l}$}
\vspace{1ex}

  In this paper we are not concerned so much with the numerical value of
$\svev{\f_0}$, but with the fact that this condensate violates the
SUSY indentity \seeq{m0}. Since we are dealing with a matter of principle,
it is worthwhile to give an alternative derivation of the existence of
SUSY violations.

  Examining the various leading order supersymmetric results found in the
literature~\cite{ads,itep} we observe that, typically, there are diagrams that
contain a different number of insertions of the Higgs field. Each classical
field provides a factor of $v$ and, to match dimensions, the diagram must
contain compensating powers of $\r$. The relative weight of different diagrams
is therefore $\r$-dependent, and the supersymmetric result depends crucially on
performing the $\r$-integration.

  In the calculation of $\svev{\f_0}$ the two diagrams Figs.~1 and~2 have a
different $\r$-dependence too. But now the $\r$-integration gives rise to a
SUSY violating result. Introducing to the dimensionless variable
\beq
  s\equiv 4\p^2 v^2 \r^2 \,,
\eeq
we can rewrite \seeq{sum} as
\beq
  \svev{\f_0} = m_0^* {\L^4\over v^4} {y^2\over 4\p^2 g^4 h} \log y
  \int ds\, e^{-s} s^2 (c_1 s - c_2) \,.
\label{sums}
\eeq
The numerical values of the constants are $2c_1=c_2=1$. The reason why we have
introduced them will be explained shortly.

  Let us now discuss the $m_0^*$-dependence of $\svev{\l\l}$.
The diagrams that contribute to the leading log are closely related
to Figs.~1 and~2. We only have to replace the $\f_0$ tadpole (Fig.~3)
by a new sub-graph in which the two $\l^{SS}$ zero modes go
directly to the external point (see Fig.~4).
The other parts of the diagrams as well as the symmetry factors are unchanged.
In the case of the diagram related to Fig.~1,
this amounts to replacing the $g^2\r^2/(8h)$ factor in \seeq{intg} by 2.
The coefficient of the leading log of $\svev{\l\l}$ takes the form
\beq
  \svev{\l\l} = m_0^* {\L^4\over v^2} {16 y^2\over g^6 } \log y
  \int ds\, e^{-s} s (c_1 s - c_2) \,.
\label{ll}
\eeq

  Notice that the expression in parenthesis is the same as in
\seeq{sums}. This expression corresponds to the parts of Figs.~1 and~2 that
are unchanged in the calculation of $\svev{\l\l}$. A comparison of
\seeqs{sums} and\seneq{ll} reveals that the integrand of \seeq{sums}
contains an extra power of $s$, \ie an extra power of $\r^2$.
Consequently, regardless of the numerical values of the
constants $c_1$ and $c_2$, {\it the leading SUSY violating logs cannot vanish
simultaneously} for $\svev{\f_0}$ and $\svev{\l\l}$.

  As we have explained above, the power of $\r^2$ in every graph is determined
by the discrepancy between the explicit factors of $v$ and the dimensionality
of the diagram (taking into account the common
dimensionful constant $m_0^* \L^4$). In the case of $\svev{\l\l}$, the
integration over the pair of $\l^{SS}$ zero modes gives rise to no
factors of $v$. In the case of the $\f_0$ tadpole, there is a factor of
$v^2$ coming from the two insertions of the Higgs field indicated explicitly
in Fig.~3. But this $v^2$ is cancelled by a $1/v^2$ coming from
the zero momentum Fourier transform of the $\f_0$ propagator:
$m^{-2}=(2 h^2 v^2)^{-1}$. This non-analytic dependence on $v^2$ arises
because the massive propagator is obtained
from the massless one by an infinite sum over an arbitrary number of mass
insertions. The overall power of $v$ is eventually the same in both cases.
Since the dimension of $\svev{\f_0}$ is smaller than the dimension
of $\svev{\l\l}$ by two, there has to be an extra power of $\r^2$ in the case
of $\svev{\f_0}$.

  Substituting the numerical values of $c_1$ and $c_2$
and performing the $\r$-integration, we find
that the coefficient of the leading log cancels between
the two terms in \seeq{ll}. Thus, contrary to the case of $\svev{\f_0}$,
there is no logarithmic enhancement for $\svev{\l\l}$.
A calculation of the contribution with no logarithmic enhancement to
$\svev{\l\l}$ should involve the following ingredients.  First, the
cancellation of logarithmic IR divergences between the two diagrams discussed
above leaves behind a finite non-zero remainder. Additional contributions
arise from massless diagrams in which pairs of classical
fields have been replaced by propagators. These diagrams are individually
finite.

  All the diagrams mentioned above belong to the same order in the massless
expansion. But this is not the end of the calculation. Comparable contributions
arise from higher order terms in the Born series. Although diagrams with more
insertions of the Higgs field are formally of higher order, they
contain stronger IR divergences that compensate for the extra powers of
the coupling constant $y$. We first observe that even terms in the Born series
represent corrections to the $\h$-component of the ``lepton'' zero modes,
whereas odd terms are corrections to the $\bar\x$-component.
Denoting the $(2n-1)$-st correction by $\bar\x^{(n)}(x)$ one has
\beq
  \bar\x^{(n)}(x) \sim \r\, m_1^{2n-1} x^{2n-4}\,, \quad\quad
  \r \ll |x| \ll m_1^{-1} \,.
\eeq
In this equation we have shown only the leading power law behaviour
and suppressed logarithmic factors. We see that the Born series represent an
expansion in powers of $m_1 x$. But the physical cutoff of the $x$-integration
occurs at $x \sim m_1^{-1}$. Thus,
after the $x$-integration each extra power of $x^2$ leads to an extra power
of $1/m_1^2$. A similar pattern is found for higher order corrections to the
induced scalar fields (the sub-graph in the upper-left part of Fig.~2).
The final result is that the massless expansion breaks down
because it becomes an expansion in $m_1^2/m_1^2=1$.

  The conclusion is that, in order to calculate the leading
order value of $\svev{\l\l}$, one must sum contributions from all orders
in the massless expansion. Within the massive Feynman rules,
this is equivalent to a calculation of the
non-logarithmic terms that arise from the exact form of the
zero modes and the induced scalar fields in the ``lepton'' sector.
We expect that eventually $\svev{\l\l}$ will be non-zero because the
linearized massive field equations are manifestly not supersymmetric.

\vspace{5ex}
\noindent {\large\bf 4.~~Discussion}
\vspace{3ex}

  The definition of the path integral measure requires one to specify
a complete set of quantum modes. In the vacuum sector, the free (massive
or massless) field equations define the same set of plane waves for both
bosons and fermions. Consequently, there are no SUSY violations in the
vacuum sector.

  In the instanton sector, there are {\it three} relevant complete
sets of modes. The first complete set consists of the massless modes pertaining
to the self-dual classical instanton solution. This basis is universal
for all spins, and it defines the formally supersymmetric perturbative
expansion. The massive field equations, that contain the Higgs field,
define two more complete sets: one for the bosons and one for the fermions.
The massive bosonic and fermionic modes are different from each other,
as well as from the massless modes.
{\it The path integral measure must be defined using the basis of massive
modes}, because the alternative basis of massless modes leads
to IR divergences.

  In fact, the massless spectrum is not exactly supersymmetric. The massless
perturbative expansion is well defined only in a finite box. Since
there are no supersymmetric boundary conditions  in the instanton sector,
the result is an $O(1/R)$ discrepancy between the bosonic and fermionic spectra
($R$ is the size of the box). The explicit $1/R$ breaking terms are
multiplied by positive powers of $R$ coming from the IR divergences, and
so they can (and do) leave behind a finite effect in the infinite volume
limit.

   The massive fluctuations spectrum, which is the appropriate one
in the infinite volume limit, is manifestly not supersymmetric. The finite
discrepancy between the spectra of massive fluctuations arises because
of the non-constancy of the Higgs field in the instanton's core.
  The crucial effect is the existence of terms that involve {\it derivatives}
of the classical Higgs field in the Schr\"odinger-like
operators that define the bosonic and fermionic modes.
(In the case of the fermions we refer to the square of the Dirac operator).
Derivatives of the Higgs field appear as potentials in these
Schr\"odinger operators, and one can explicitly check that the potentials
are different for bosons and fermions for any choice of the Higgs field.

  Let us now reexamine the SUSY violating result found in Sect.~3.
The existence of different powers of $\r$ in the final expressions
\seeqs{sums} and\seneq{ll} means that the saddle point of the $\r$-integration
is different in each diagram. In other words, the effective value of $\r$
depends on the operator that we measure. This is not surprising.
The integration over $\r$ is gaussian, and so $\r$ really represents a
quantum mode.

  The logarithmic terms arise from the product of two wave functions that
behave like $1/x^2$ in the intermediate region. They depend on the
effective value of $\r$ in two ways. The most important effect is that the
normalization of the relevant wave functions is $\r$-dependent. A secondary
effect is that the short distance cutoff of the $1/x^2$ behaviour
is determined by $\r$. In the intermediate region, the equations
of motion are locally supersymmetric to a first approximation.
The SUSY violation arises because one cannot solve the equations with
supersymmetric boundary conditions both a the origin and at infinity.
Physically, the asymptotic region is supersymmetric because it is
a free vacuum. The breaking of SUSY originates from the
Higgs dependent part of instanton's core.
In the calculation, it takes the form of non-supersymmetric values for
the effective $\r$.

  The Higgs model and the observables discussed in this
paper were chosen for reasons of technical convenience. In essence,
the anomalous breaking of SUSY was shown to be directly related to
the IR divergences of massless perturbation theory. We thus conjecture
that the SUSY anomaly is a generic property of asymptotically free
supersymmetric gauge theories.

  Specifically, we  proved the failure of the holomorphicity of physical
observables constructed from the lowest components of gauge invariant chiral
superfield. The conventional treatment of supersymmetric QCD depends
heavily on the assumed holomorphicity, leading to conclusions such as
the existence of a pathological run-away behaviour. We believe that
these conclusions have to be reexamined in view of the failure of
holomorphicity found in this paper.

\vspace{5ex}
\centerline{\rule{5cm}{.3mm}}


\newpage
\noindent {\large\bf Appendix}
\secteq{A}

\vspace{2ex}
\noindent {\bf I.~~Zero modes in the pure instanton background}
\vspace{1ex}

\noindent  In the instanton sector, the classical gauge field is given by
\beq
  A^a_\m = {2\over g} {\bar\h_{a\m\n}\, x^\m \over r }\, a(r)\,,
\label{a}
\eeq
and the classical Higgs field is
\beq
  \f_{iA} = i v\, {\bar\s^\m_{iA}x^\m \over r }\, \varphi(r) \,.
\label{hf}
\eeq
We have suppressed the collective coordinate $x_0^\m$.
For $mr \ll 1$ one has
\beq
  a(r) = r/(r^2+\r^2) \,,
\label{ax}
\eeq
\beq
  \varphi(r)= r/(r^2+\r^2)^{{1\over 2}} \,.
\label{fx}
\eeq
For $mr \gg 1$ these expressions are no longer valid. Both the gauge field
and the Higgs field tend exponentially to their limiting values
$a(r)\to 1/r$ and $\varphi(r)\to 1$. In this paper we only need
the explicit form of the classical fields for $mr  \ll 1$.

  We next turn to the fermionic zero modes. The measure \seeq{msr} requires
the normalization $\int d^4x\, |\J|^2 =1$ for all zero modes.
In the pure instanton background the normalized gaugino zero modes
are\footnote{
  The euclidean partition function of supersymmetric theories is defined
using a Majorana representation for the fermions. Consistency of this
representation determines the global phase of the
zero modes' wave functions. The Majorana representation
is mandatory for the gaugino but it is convenient to treat the matter
fermions analogously. See \eg ref.~\cite{cs1} for more details.
}
\bqry
   (\l^a_\a)^{SS}_k & = & {\sqrt{2}\,\r^2\over \p}
                           {\s^a_{\a k} \over (x^2+\r^2)^2 }\,,
\label{lss} \\
   (\l^a_\a)^{SC}_k & = & {\r \over \p}
        {\s^a_{\a l} x^\m \s^\m_{lk} \over (x^2+\r^2)^2 }\,,
\label{lsc}
\eqry
where $k=1,2$. The zero modes of the four charged doublets are identical
except for the flavour quantum numbers. The Higgsino zero modes are
\beq
  (\j_{iA\a})_n = {\r \over \p}
                  {\d_{in}\d_{A\a}\over (x^2+\r^2)^{3\over 2}} \,,
\label{higgsino}
\eeq
where $n=1,2$. The ``lepton'' zero modess are
\beq
  \h_{A\a\pm} = {\r \over \p}
                  {\d_{A\a}\over (x^2+\r^2)^{3\over 2}} \,.
\label{eta}
\eeq

\vspace{2ex}
\noindent {\bf II.~~Exact zero modes}
\vspace{1ex}

  When we turn on the Higgs field, the $\l^{SS}$ and the $\h_\pm$ pairs
remain exact zero mode, but with new field component and modified
wave functions. The $\l^{SS}$ pair now takes the form~\cite{ads}
\beqrabc{A.9}
   (\l^a_\a)_k & = & \s^a_{\a k}\, \hat{f}
\label{gg_a} \\
   (\bar\j_{iA\a})_k & = & i \e_{i\a} \d_{Ak}\, \hat{g}
                +i \e_{\b\a} x^\m x^\n \s^\m_{Ai} \bar\s^\n_{\b k}\, \hat{h}
\label{gg_b} \\
   (\j_{0\a})_k & = & i \d_{\a k}\, \hat{p}
\label{gg_c}
\eqry
The hat denotes radial functions. The ``lepton'' zero modes are
now given by
\beqrabc{A.10}
   \h_{A\a\pm} & = &  \d_{\a A}\, \hat{u}
\label{lept_a} \\
   \bar\x_{i\a\mp} & = & \pm \e_{i\a}\, \hat{w}
\label{lept_b}
\eqry
\renewcommand{\theequation}{A.\arabic{equation}}
\setcounter{equation}{10}
\noindent The quantum numbers of the different field components of
these zero modes
can be found in Table~2. The conserved angular momenta in the instanton
sector are
\bqry
  K^a_1 & = & S^a_1 + L^a_1 + T^a \,, \NON
  K^a_2 & = & S^a_2 + L^a_2 + F^a \,.
\eqry


\vspace{3ex}

%
%
\begin{table}[htb]
\begin{center}
\begin{tabular}{|c|c c c c c c c|}
\hline
channel & $S_{1}$ & $S_{2}$ & T & F & L & $K_{1}$ & $K_{2}$
\\ \hline \hline
$\l^a$ & $\frac{1}{2}$ & 0 & 1 & 0 & 0 & $\frac{1}{2}$ & 0
\\ \hline
$\bar\j_{iA}$ & 0 & $\frac{1}{2}$ & $\frac{1}{2}$ &
$\frac{1}{2}$ & 0,1 & $\frac{1}{2}$ & 0
\\ \hline
$\j_0$ & $\frac{1}{2}$ & 0 & 0 & 0 & 0 & $\frac{1}{2}$ & 0
\\ \hline \hline
$\h_{A}$  & $\frac{1}{2}$ & 0 & $\frac{1}{2}$ & 0 & 0 & 0 & 0
\\ \hline
$\bar\x_{i}$  & 0 & $\frac{1}{2}$ & 0 & $\frac{1}{2}$ & 0 & 0 & 0
\\ \hline
\end{tabular}
\end{center}
\caption{
  Quantum numbers of the field components of the exact fermionic zero
  modes. }
\end{table}


The radial functions are obtained by solving ordinary coupled differential
equations. The equations for the lepton zero modes are
\beqrabc{A.12}
   \hat{u}' + 3 a\, \hat{u} & = & - m_1 \varphi\, \hat{w}
\label{slpt_a} \\
   \hat{w}' & = & - m_1 \varphi\, \hat{u}
\label{slpt_b}
\eqry
The prime denotes differentiation with respect to $r$. The functions
$a=a(r)$ and $\varphi=\varphi(r)$ are defined in \seeqs{a} and\seneq{hf}.
The equations for the $\l^{SS}$ zero modes are
\beqrabc{A.13}
  \hat{f}' + 4 a \hat{f}  & = & - (\m/\sqrt{2}) \varphi\, \hat{g}
\label{ss_a} \\
  \hat{g}' + (1/r-2a)\hat{g} + (a-1/r)\hat{h}_1
       & = & - \sqrt{2}\m\varphi \hat{f}
\label{ss_b} \\
  \hat{p}'  & = & (m/\sqrt{2}) \varphi\, \hat{h}_1
\label{ss_c} \\
  \hat{h}'_1 + (3/r)\hat{h}_1 + 3(a-1/r)\hat{g}
        & = & \sqrt{2} m\varphi\, \hat{p}
\label{ss_d}
\eqry
\renewcommand{\theequation}{A.\arabic{equation}}
\setcounter{equation}{13}
\noindent where
\beq
   \hat{h}_1 = \hat{g} + 2 r^2 \hat{h} \,.
\eeq
The mass parameters $\m$, $m$ and $m_1$ are defined in Sect.~2.
Notice that the pairs $(\hat{f},\hat{g})$
and $(\hat{p},\hat{h}_1)$ diagonalize the mass operator at infinity.

  The Born approximation amounts to neglecting the
off-diagonal term on the \rhs of \seeq{slpt_a}.
One solves \seeq{slpt_b} with $\hat{u}$
given by the pure instanton expression. (Compare \seeq{lept_a}
and \seeq{eta}). The result is \seeq{approx}. Similarly, solving
\seeqs{ss_b} and\seneq{ss_d} with $\hat{f}$ given by the pure instanton
expression (see \seeq{lss}) and with $\hat{p}=0$ gives rise to
\seeq{newj}.

  As discussed in Sect.~3,  the calculation of the leading log arising from
the ``lepton'' zero mode \seeq{intx} requires a knowledge of the scale
at which \seeq{approx} is no longer valid.  The asymptotic behaviour of the
wave function is a falling exponential
\beq
   \hat{u}=\hat{w}= c_0 \left( {m_1 \over r^3} \right)^{1\over 2}
                    e^{-m_1 r} \,.
\label{asymp}
\eeq
The transition from the power law behaviour given
by \seeqs{eta} and\seneq{approx} to the asymptotic behaviour
\seeq{asymp} occurs at the distance scale $m_1^{-1}$. This is a general
property that can be derived as follows.
Let us approximate the wave function by
\seeqs{eta} and\seneq{approx} for $r\le r_0$, and  by \seeq{asymp}
for $r\ge r_0$. We thus replace the smooth transition region by a
sharp transition at $r=r_0$.
Requiring that the two components of the wave function be continuous
across the transition point gives us two equations for the two unknowns
$r_0$ and $c_0$. The solution is $r_0=O(m_1^{-1})$ and $c_0=O(y)$.
This information is sufficient to determine the leading log in \seeq{intx}.

  The physical cutoff of the integral in \seeq{bosons}
occurs at the same scale. This is derived by applying a similar analysis
to the bosonic field equations. The same techniques
can be used to obtain the qualitative structure of the transition
region for the gaugino zero modes, but this information is not necessary
for our calculations.

\newpage
\noindent {\bf III.~~Approximate zero modes and the massive propagator}
\vspace{1ex}

 The last issue that we discuss is the approximate zero modes originating
from the Higgsino and $\l^{SC}$ zero modes, and their relation to the
propagators of the massive Feynman rules. Going to a partial waves
basis, let us denote by $G_0$ the massless propagator with the quantum
numbers of the Higgsino and $\l^{SC}$ zero modes,
\ie $K_1=0$ and $K_2={1\over 2}$.
The zero modes themselves will be denoted by the generic name $\J_0$.
The massless radial propagator $G_0$ obeys the differential equation
\beq
  H_0\, G_0(r',r) = r^{-3}\, \d(r-r') - \sket{\J_0} \sbra{\J_0} \,.
\label{proj}
\eeq
Notice the second term on the \rhs of \seeq{proj} that projects out
the zero modes. On the other hand, the massive propagator $G$
with the same quantum numbers solves the equation
\beq
  (H_0+V)\, G(r',r) = r^{-3}\, \d(r-r') \,.
\label{nop}
\eeq
Now there is a simple delta function on the \rhs of the equation,
because there are no
exact zero modes with the given quantum numbers in the presence of the Higgs
field.

  We now show that at short distances, the exact
propagator $G$ must have a ``large'' component which represents the
approximate zero modes. This will justify the use of the massless
Feynman rules for the Higgsino and $\l^{SC}$ zero modes in Figs.~1 and~2.
We first observe that the matrix element of the Higgs field
\beq
  E_1 = \sbra{\J_0} V \sket{\J_0}
\eeq
between the zero modes is small (Explicitly $E_1=\e_{kn}\, gv/\sqrt{2}$ where
the indices $k$ and $n$ count the superconformal and Higgsino zero modes
respectively). If the zero modes have turned into finite energy bound states
$\J$ with $E \sim E_1$, the large component would simply be
$\sket{\J} E^{-1} \sbra{\J}$. Now, the support of the zero modes is inside
a potential well of radius $\r$, and so for $r\ll m^{-1}$ it should make
no difference whether the zero modes have turned into true bound states or into
resonances. Let us denote the first Born correction by
\beq
  \J_1 = - G_0 V \J_0 \,.
\eeq
One can check by direct substitution that for $r\ll m^{-1}$, the massive
propagator $G$ is related to the massless propagator $G_0$ by
\beq
  G = G_0 + \sket{\J_0+\J_1} E_1^{-1} \sbra{\J_0+\J_1} + O(g) \,.
\label{rlate}
\eeq
In the instanton's core, the \rhs of \seeq{rlate} contains an $O(1/g)$ term
which is an antisymmetric product of the original zero modes.

  Within the massive Feynman rules, the first factor in \seeq{intg} is regarded
as a part of fermionic determinant, \ie as a part of the measure.
Keeping track of the relation
between the diagrams drawn using the massive and the massless Feynman rules,
we find that the application of the massless Feynman rules to the
Higgsino and $\l^{SC}$ zero modes in Figs.~1 and~2 is justified.

  Last we discuss the cancellation of logarithmic IR divergences
related to the Higgsino and $\l^{SC}$ zero modes. Such a logarithmic
divergence can arise within the massless rules, if the two $\bar\j(x)$
wave functions which are to be integrated over at the
$\bar\j\bar\j\f_0^*$ vertex behave like $1/x^2$ for $\r\ll |x|$.
As an example, the integration over that vertex is logarithmically divergent
in each of the three diagrams shown in Fig.~5. But if we first sum
the three integrands (still using the massless Feynman rules) the
integration over the $\bar\j\bar\j\f_0^*$ vertex becomes IR convergent.
(As in Sect.~3, the integration over that vertex is done after the
integration over the external point \seeq{intprop}. Notice also that the
diagrams in Fig.~5 contain no logarithmic factors due to the $\bar\x\bar\x$
vertex. The propagator lines, which replace the classical fields in Fig.~1,
lead to a $\bar\x(x)$ wave function which is damped at least by
an extra $1/x^2$).

  In Fig.~5(a), the logarithmic divergence arises from the first
Born correction to the $\l^{SC}$ zero mode, which behaves like
$\bar\j(r)\sim 1/r^2$ for $r\gg\r$. Fig.~5(b) is a mixed case.
In Fig.~5(c), the logarithmic divergence arises
from the partial wave of the $G_0=\svev{\j\,\bar\j}_0$ propagator with the same
total angular momenta as the Higgsino and $\l^{SC}$ zero modes.
For $r,r'\gg\r$, the propagator $G_0(r',r)$ has a piece that behaves like
$1/(r^2 \, r'^3)$. The $\bar\j$ end of the propagator behaves like $1/r^2$.
(This is most easily seen by writing the propagator in a Dirac spinor basis).
The reason for this unusual behaviour is the projection on the (Higgsino)
zero modes on the \rhs of \seeq{proj}.

  Notice that the product of a Higgsino zero mode $\j(r')$ times the
first Born correction $\bar\j(r)$ to a $\l^{SC}$ zero mode
behaves like $1/(r^2 \, r'^3)$ too. An inspection of \seeq{rlate} reveals that
the massive radial propagator $G(r',r)$ contains the two sources of a
$1/(r^2 \, r'^3)$ behaviour described above.

  We will now show that the two $1/(r^2 \, r'^3)$ contributions on the \rhs
of \seeq{rlate} exactly cancel each other .
The massive propagator $G(r',r)$ is a solution of \seeq{nop}.
In the intermediate region $m^{-1}\gg r,r'\gg\r$,
one can neglect {\it both} the gauge field and the Higgs field.
The dominant term in the equation is the free massless Dirac operator.
Every piece of the radial propagator $G(r',r)$ must therefore be the product
of two homogeneous solutions of the free massless Dirac equation.
For $r>r'$, the possible terms are $1/r^3$ and $1/(r\,r')^3$.
Corrections to these terms are damped by inverse powers of $r$ and/or $r'$.
In particular, a $1/(r^2 \, r'^3)$ behaviour is inconsistent with \seeq{nop}
for $r>r'$, and so it must cancel between the first and second terms on
the \rhs of \seeq{rlate}.

  Within the massive Feynman rules, the three diagrams of the massless
Feynman rules (Fig.~5) are contained in a single diagram (Fig.~6).
In this diagram one should use massive propagators. The thick lines now
correspond to various components of {\it exact} zero modes.
Sub-graphs describing the mixing of the Higgsino and $\l^{SC}$ zero modes
are absent, as they are regarded as a part of the the fermionic determinant.
All other effects of the approximate zero modes are now contained in the
massive fermion propagators.

  The cancellation of the $1/(r^2 \, r'^3)$ terms on the \rhs of \seeq{rlate}
implies that the massive diagram Fig.~6 makes no logarithmic contribution
to $\svev{\f_0}$. We comment that, as a consequence of the absence of
logarithmic terms, the sum of the three massless diagrams (Fig.~5) will be
finite provided we amputate the external $\f_0$ leg. Following the general
pattern discussed in the introduction, that sum will be equal to the
corresponding amputated massive diagram up to higher order corrections.

  As we have explained in Sect.~3, this
cancellation occurs because subleading corrections to the wave
function of an approximate zero mode cannot be disentangled from the
continuous spectrum. This behaviour is natural if the zero mode
has turned into a resonance. The new $\bar\j$ component of the $\l^{SC}$ zero
mode is then responsible for its ultimate decay.
In principle, there is also the possibility that the
approximate zero mode has turned into a finite energy bound state.
But since the energy of such a bound state is $O(m)$, it
should behave like a resonance in the intermediate region
$m^{-1}\gg r \gg \r$. The $\bar\j$ component of the $\l^{SC}$ zero mode
should then starts decaying outside of the potential barrier of radius $\r$,
but eventually it would remain trapped due to the mass term that changes the
asymptotic value of the potential.

  The massless diagrams shown in Fig.~5 involve the same $m_0^*$-dependent
vertex as Fig.~1. There is a similar set of massless diagrams with the
$m_0^*$-dependent vertex of Fig~2. Their sum is contained in the massive
diagram shown in Fig.~7. (These massless diagrams
can be retrieved from Fig.~7 in the same way that the massless diagrams of
Fig.~5 are retrieved from Fig.~6). As before, the massive diagram Fig.~7
makes no logarithmic contribution to $\svev{\f_0}$. Finally, in Fig.~8 we
show two massive diagrams that contain a closed loop. Their sum
is UV finite, and neither diagram contains a logarithmic IR term.

\vspace{5ex}
\centerline{\rule{5cm}{.3mm}}




\newpage
\noindent {\large\bf Figure captions}
\vspace{3ex}

\begin{enumerate}

\item One of the two leading log contributions to $\svev{\f_0}$. The diagrams
in Fig.~1 up to Fig.~5 are drawn using the massless Feynman rules.
The thick lines that emanate from the instanton (the shaded
circle) represent fermionic zero modes. The dotted lines that end with a cross
represent insertions of the classical Higgs field.
The vertex marked with a thick cross is linear in $m_0^*$.

\item The other leading log contribution to $\svev{\f_0}$.

\item The $\f_0$ tadpole: the common lower part of Figs.~1 and~2.

\item In the calculation of $\svev{\l\l}$ this sub-graph replaces the
$\f_0$ tadpole in Figs.~1 and~2.

\item Cancellation of logarithmic IR divergences related to the superconformal
and Higgsino zero modes. In each diagram the integration over the
$\bar\j\bar\j\f_0^*$ vertex is logarithmically IR divergent.
As shown in part III of the Appendix, the logarithmic IR divergences cancel
among these diagrams before the  $\r$-integration.

\item This diagram, drawn using the massive Feynman rules, corresponds to the
sum of the three massless diagrams in Fig.~5. It makes no logarithmic
contribution to $\svev{\f_0}$.

\item Another massive diagram that makes no logarithmic
contribution to $\svev{\f_0}$. This diagram contains the same
$m_0^*$-dependent vertex as Fig.~2.

\item Two massive diagrams containing a closed loop. Their sum is UV finite,
and neither diagram contains a logarithmic IR term.

\end{enumerate}

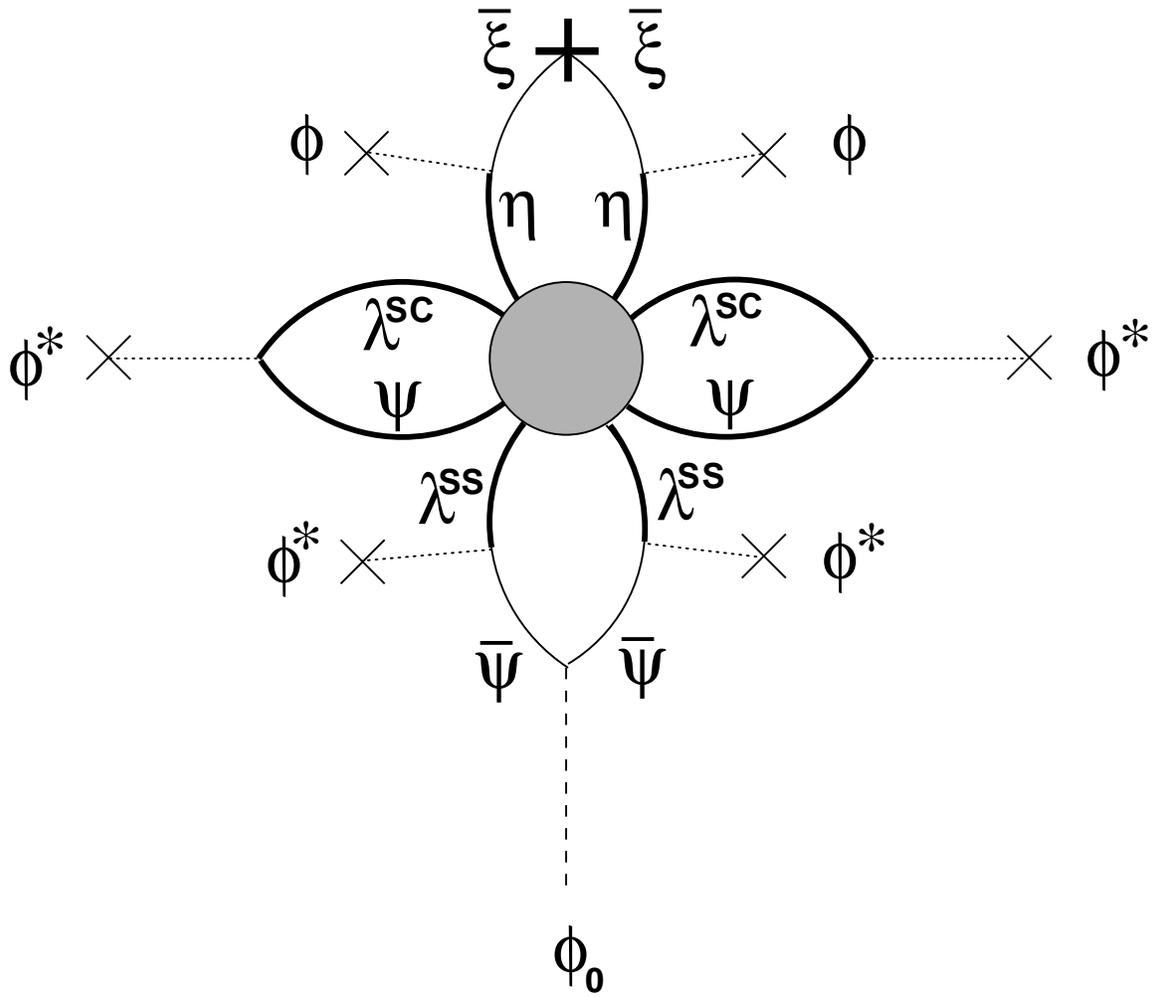
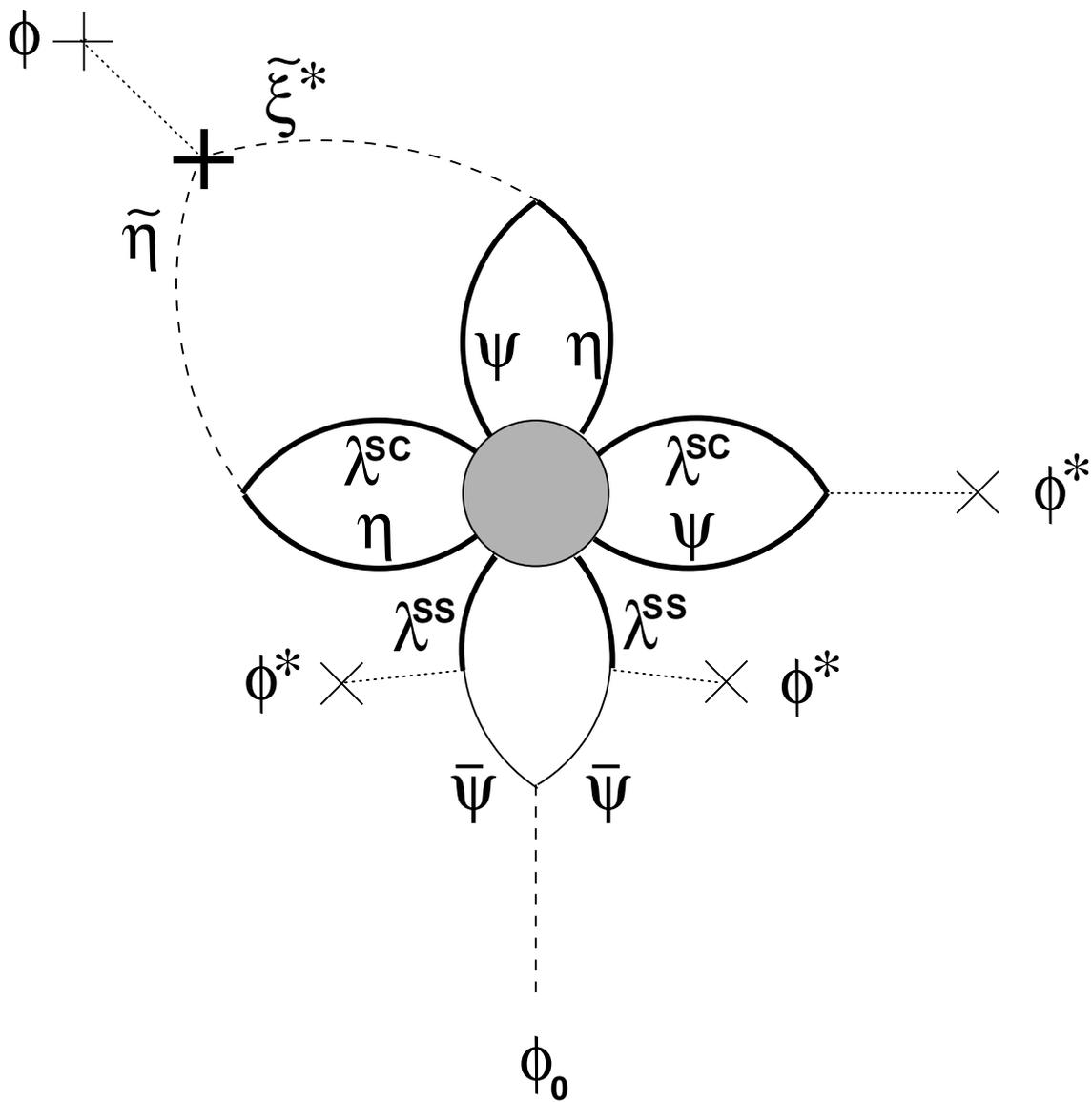
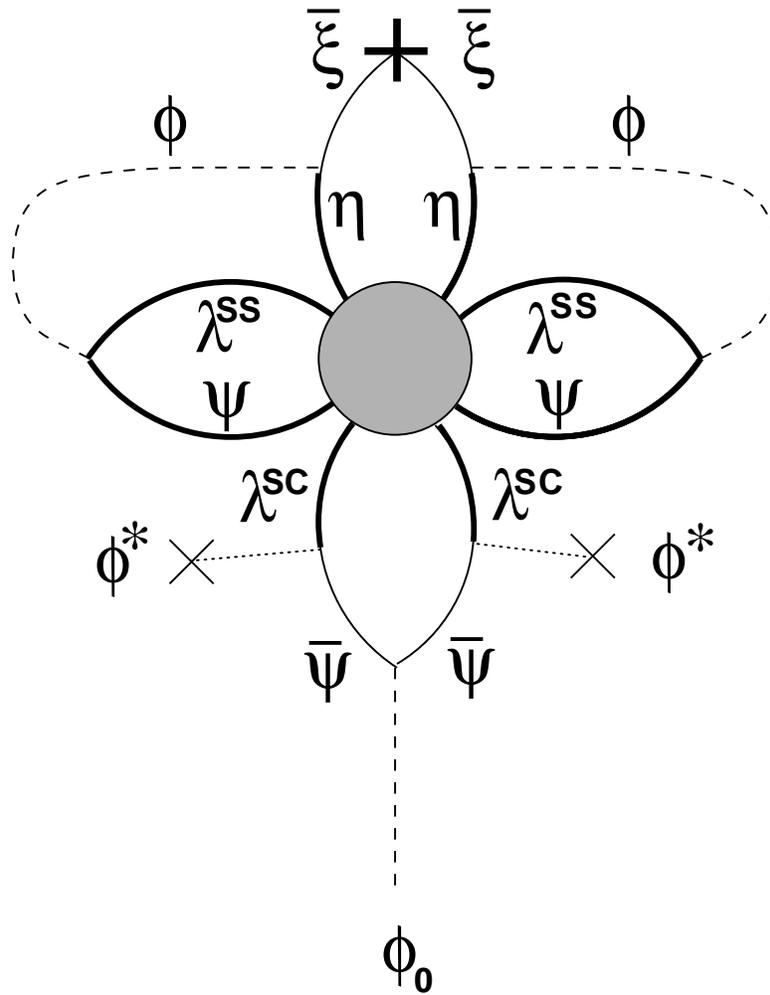
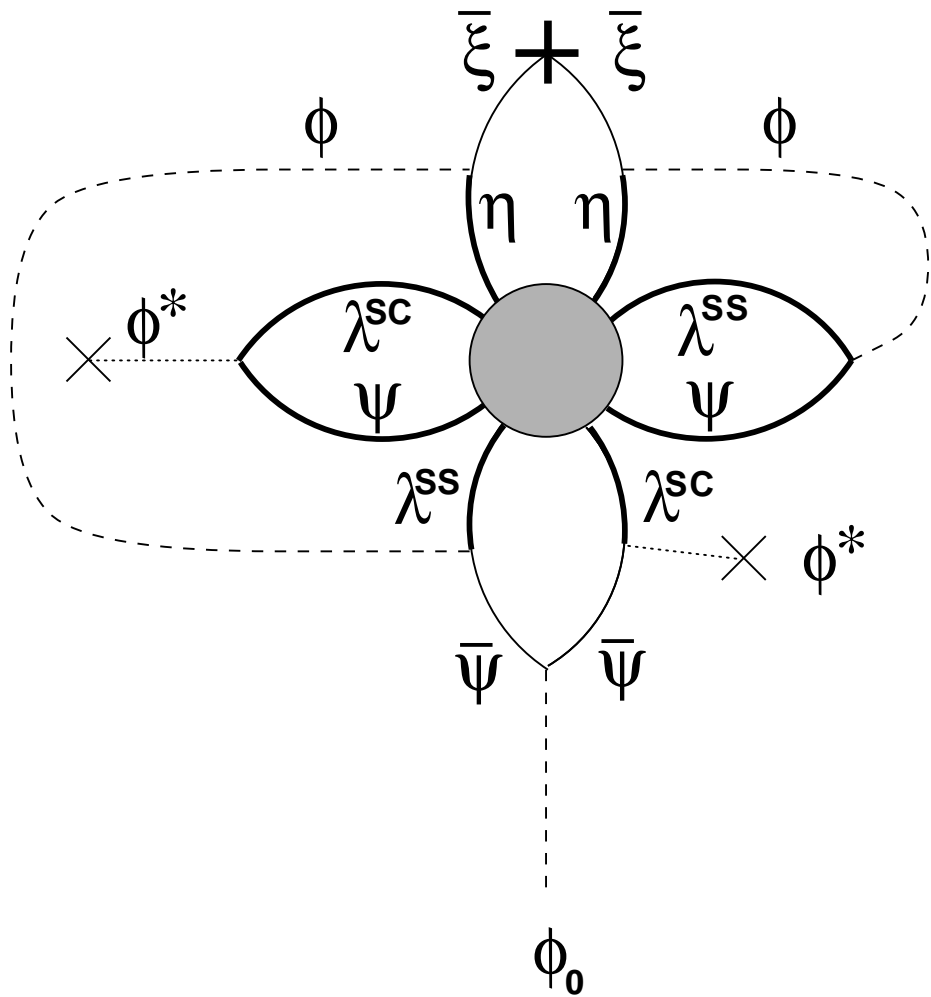
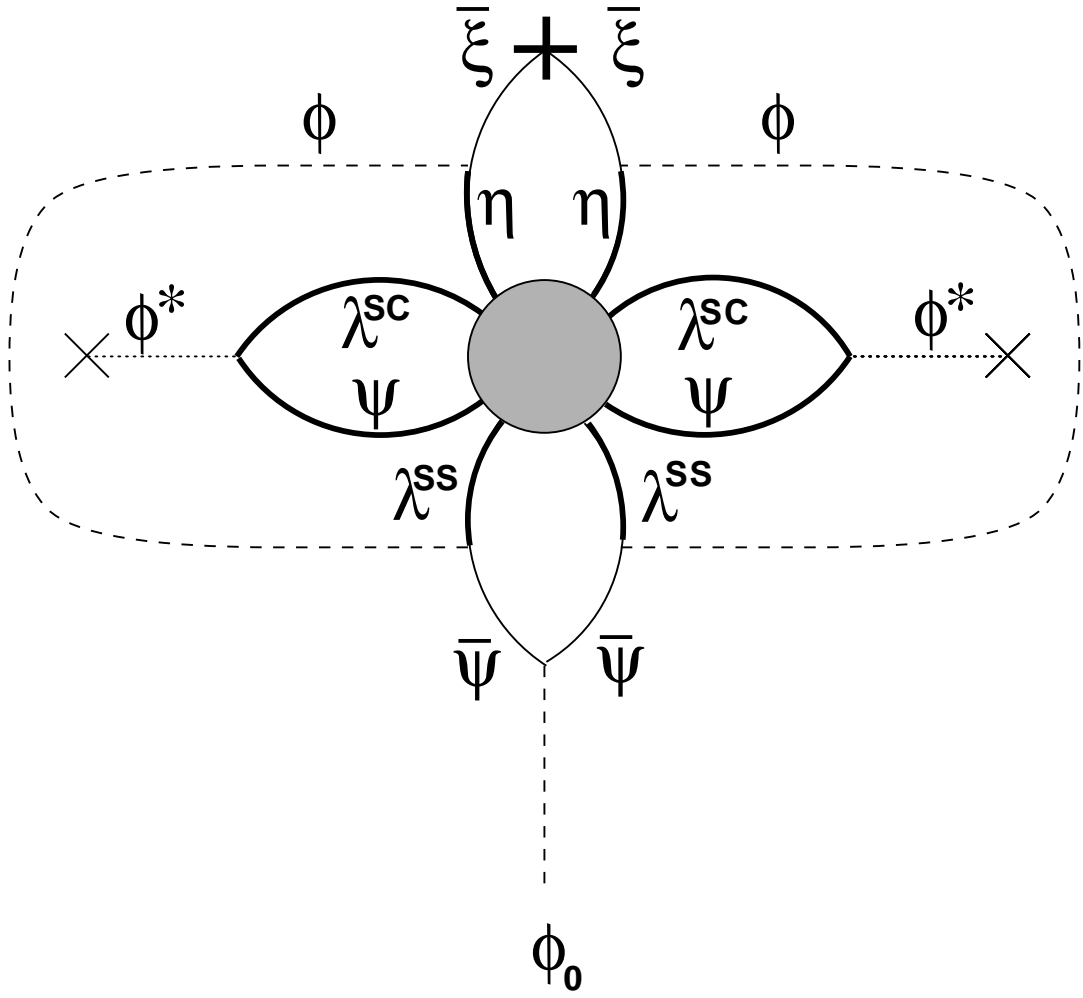
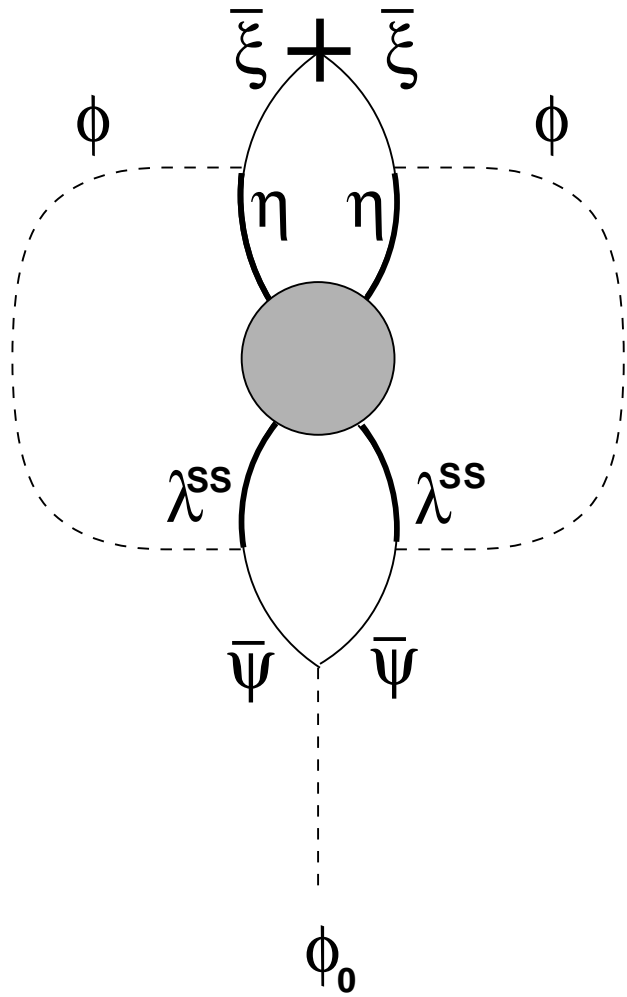
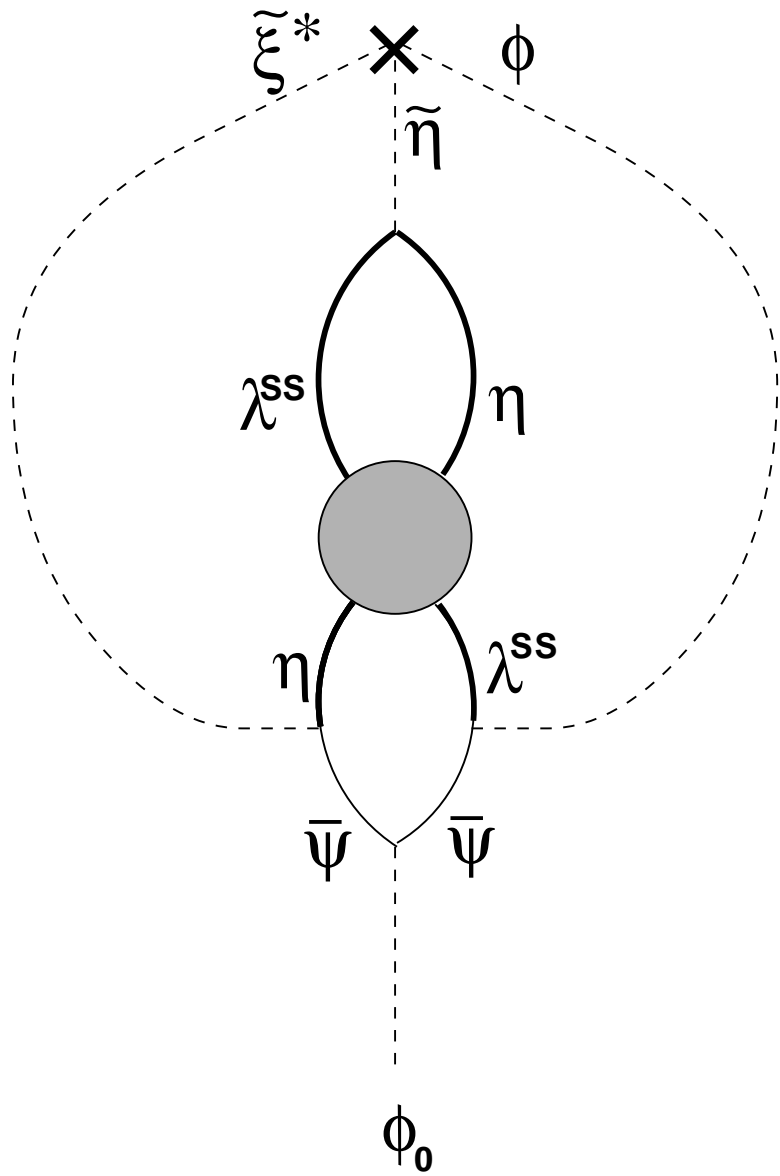
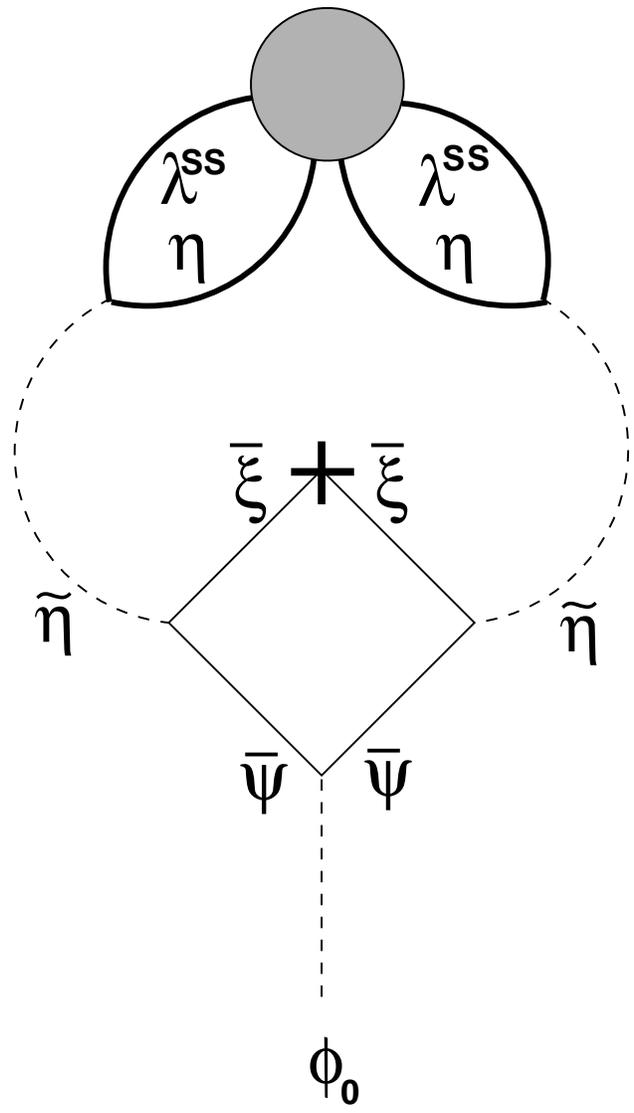
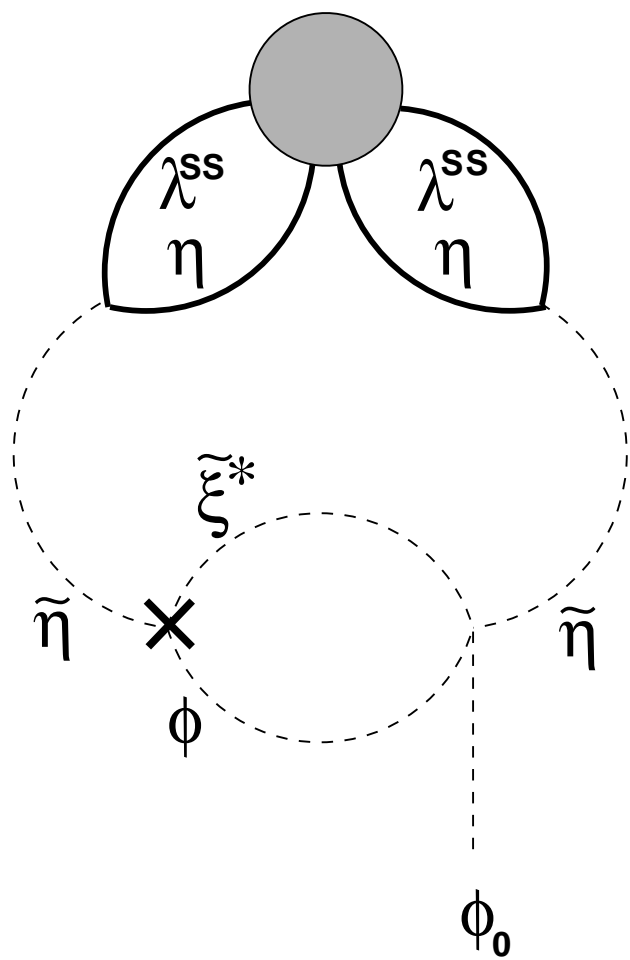
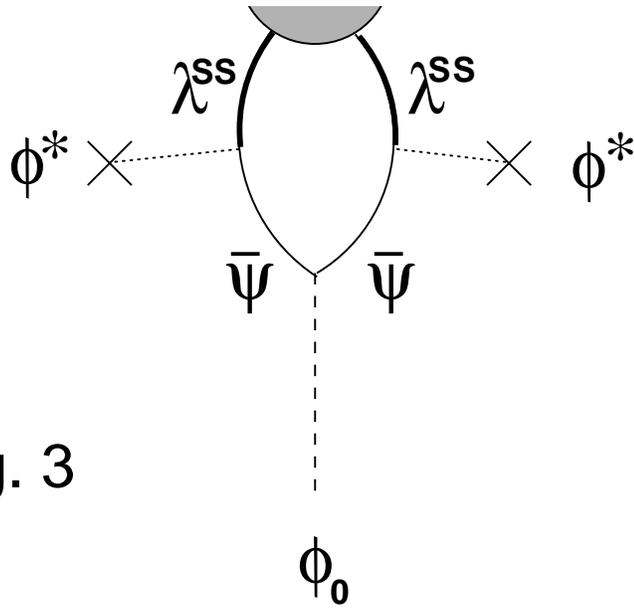
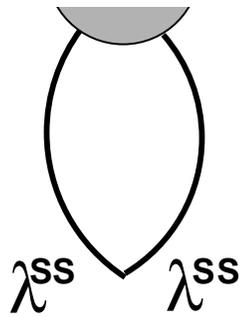
\end{document}